\documentclass[article, aps, 12pt, floatfix, longbibliography, a4paper]{revtex4-2}

\usepackage{graphicx}
\usepackage{epstopdf}
\usepackage{amssymb}
\usepackage{hyperref}

\usepackage{color}
\usepackage{amsmath,bm}
\usepackage[utf8]{inputenc}
\usepackage{xcolor}
\usepackage{physics}

\usepackage{placeins}

\newcounter{hoge}
\makeatletter
\@addtoreset{figure}{hoge}
\makeatother
\makeatletter
\@addtoreset{equation}{hoge}
\makeatother
\makeatletter
\@addtoreset{section}{hoge}
\makeatother
\makeatletter
\@addtoreset{table}{hoge}
\makeatother
\makeatletter
\@addtoreset{equation}{hoge}
\makeatother

\begin{document}

\title[]{Perfectly harmonic spin cycloid and multi-$Q$\\
textures in the Weyl semimetal GdAlSi}


\author{Ryota Nakano$^{1}$}
\author{Rinsuke Yamada$^{1}$}\email{ryamada@ap.t.u-tokyo.ac.jp}
\author{Juba Bouaziz$^{2}$}\email{jbouaziz@g.ecc.u-tokyo.ac.jp}
\author{Maurice Colling$^{3,\P}$}
\author{Masaki Gen$^{4,5}$}
\author{Kentaro Shoriki$^{1}$}
\author{Yoshihiro Okamura$^{1}$}
\author{Akiko Kikkawa$^{4}$}
\author{Hiroyuki Ohsumi$^{6}$}
\author{Yoshikazu Tanaka$^{6}$}
\author{Hajime Sagayama$^{7}$}
\author{Hironori Nakao$^{7}$}
\author{Yasujiro Taguchi$^{4}$}
\author{Youtarou Takahashi$^{1,4}$}
\author{Masashi Tokunaga$^{2,3}$}
\author{Taka-hisa Arima$^{4,8}$}
\author{Yoshinori Tokura$^{1,4,9}$}
\author{Ryotaro Arita$^{5,10}$}
\author{Jan Masell$^{3,4}$}
\author{Satoru Hayami$^{11}$}
\author{Max Hirschberger$^{1,4}$}\email{hirschberger@ap.t.u-tokyo.ac.jp}

\affiliation{$^{1}$Department of Applied Physics and Quantum-Phase Electronics Center, The University of Tokyo, 113-8656, Tokyo, Japan}
\affiliation{$^{2}$Research Center for Advanced Science and Technology, The University of Tokyo, Tokyo 153-8904, Japan}
\affiliation{$^{3}$Institute of Theoretical Solid State Physics, Karlsruhe Institute of Technology (KIT), 76049 Karlsruhe, Germany}
\affiliation{$^{4}$RIKEN Center for Emergent Matter Science (CEMS), Wako, Saitama 351-0198, Japan}
\affiliation{$^{5}$Institute for Solid State Physics, The University of Tokyo, Kashiwa 277-8581, Japan}
\affiliation{$^{6}$RIKEN SPring-8 Center, 1-1-1 Kouto, Sayo, Hyogo 679-5148, Japan}
\affiliation{$^{7}$Institute of Materials Structure Science, High Energy Accelerator Research Organization, Tsukuba, Ibaraki 305-0801, Japan}
\affiliation{$^{8}$Department of Advanced Materials Science, The University of Tokyo, Kashiwa 277-8561, Japan}
\affiliation{$^{9}$Tokyo College, The University of Tokyo, Tokyo 113-8656, Japan}
\affiliation{$^{10}$Department of Physics, The University of Tokyo, Tokyo, 113-0033, Japan}
\affiliation{$^{11}$Graduate School of Science, Hokkaido University, Sapporo 060-0810, Japan}
\affiliation{$^{\P}$Current address: Department of Materials Science and Engineering, Norwegian University of Science and Technology (NTNU), Trondheim 7491, Norway}


\maketitle
\newpage


\begin{center}
\Large{Abstract}
\end{center}

A fundamental question concerns how topological electronic states are influenced by many-body correlations, and magnetic Weyl semimetals represent an important material platform to address this problem.
However, the magnetic structures realized in these materials are limited, and in particular, no clear example of an undistorted helimagnetic state has been definitively identified.
Here, we report clear evidence of a harmonic helimagnetic cycloid with an incommensurate magnetic propagation vector in the Weyl semimetal GdAlSi via resonant elastic X-ray scattering, including rigorous polarization analysis. This cycloidal structure is consistent with the Dzyaloshinskii-Moriya (DM) interaction prescribed by the polar crystal structure of GdAlSi. Upon applying a magnetic field, the cycloid undergoes a transition to a novel multi-$Q$ state. This field-induced, noncoplanar texture is consistent with our numerical spin model, which incorporates the DM interaction and, crucially, anisotropic exchange. \textit{The perfectly harmonic Weyl helimagnet GdAlSi} serves as a prototypical platform to study electronic correlation effects in periodically modulated Weyl semimetals.

\newpage


\begin{center}
\Large{Main Text}
\end{center}

\textbf{Introduction.}\\
Inversion-breaking Weyl semimetals host two-fold degenerate band crossings with a quasi-relativistic, linear energy-momentum dispersion and protected surface Fermi arcs~\cite{Armitage2018}. 
As such, inversion breaking Weyl semimetals provide a platform to study the interplay between many-body correlations and topologically nontrivial electronic states. Relevant proposals include the formation of an axion insulator~\cite{Wang2013,Bitan2015} and a three-dimensional fractional quantum Hall effect (3D FQHE) from disordered domains in a Weyl charge-density-wave~\cite{Sehayek2020, Yi2023, Yi2025}. Such states are realized by band-folding of Weyl fermions into a small Brillouin zone upon formation of periodically modulated order, and conditions for the hybridization of Weyl points have been derived and phrased in the language of special relativity~\cite{Chiu2023}. In this context, not only charge order but also the formation of periodically modulated helimagnetic states in inversion broken Weyl semimetals, and the question of the relevant magnetic interactions in such Weyl semimetals, has been a topic of interest~\cite{Chang2015, Hosseini2015,Araki2016, Hirschberger2021, gaudet_2021_NdAlSi, yao_2023_SmAlSi}. However, an ideal platform for studying the interplay of Weyl fermions with helimagnetism -- a perfect and undistorted helimagnetic structure in an inversion breaking Weyl semimetal -- has never been reported by quantitative polarization analysis.

Here, we report the undistorted, cycloidal helimagnetic ground state of GdAlSi with perfectly isotropic Gd$^{3+}$ ions; the experiment supersedes a prior numerical prediction of collinear antiferromagnetism (altermagnetism) in this material~\cite{Nag_2024_GdAlSi,laha_2024_GdAlSi}. We further demonstrate a field-induced transition to a complex staggered cone structure, which represents a superposition of two magnetic waves, i.e., a multi-$\bm{Q}$ state. Our numerical spin calculations successfully reproduce the ground state and field-induced transitions of GdAlSi, emphasizing the role of symmetry-allowed Dzyaloshinskii-Moriya (DM) and anisotropic exchange interactions.
\\

\begin{figure}[h]%
\centering
\includegraphics[width=1.0\textwidth]{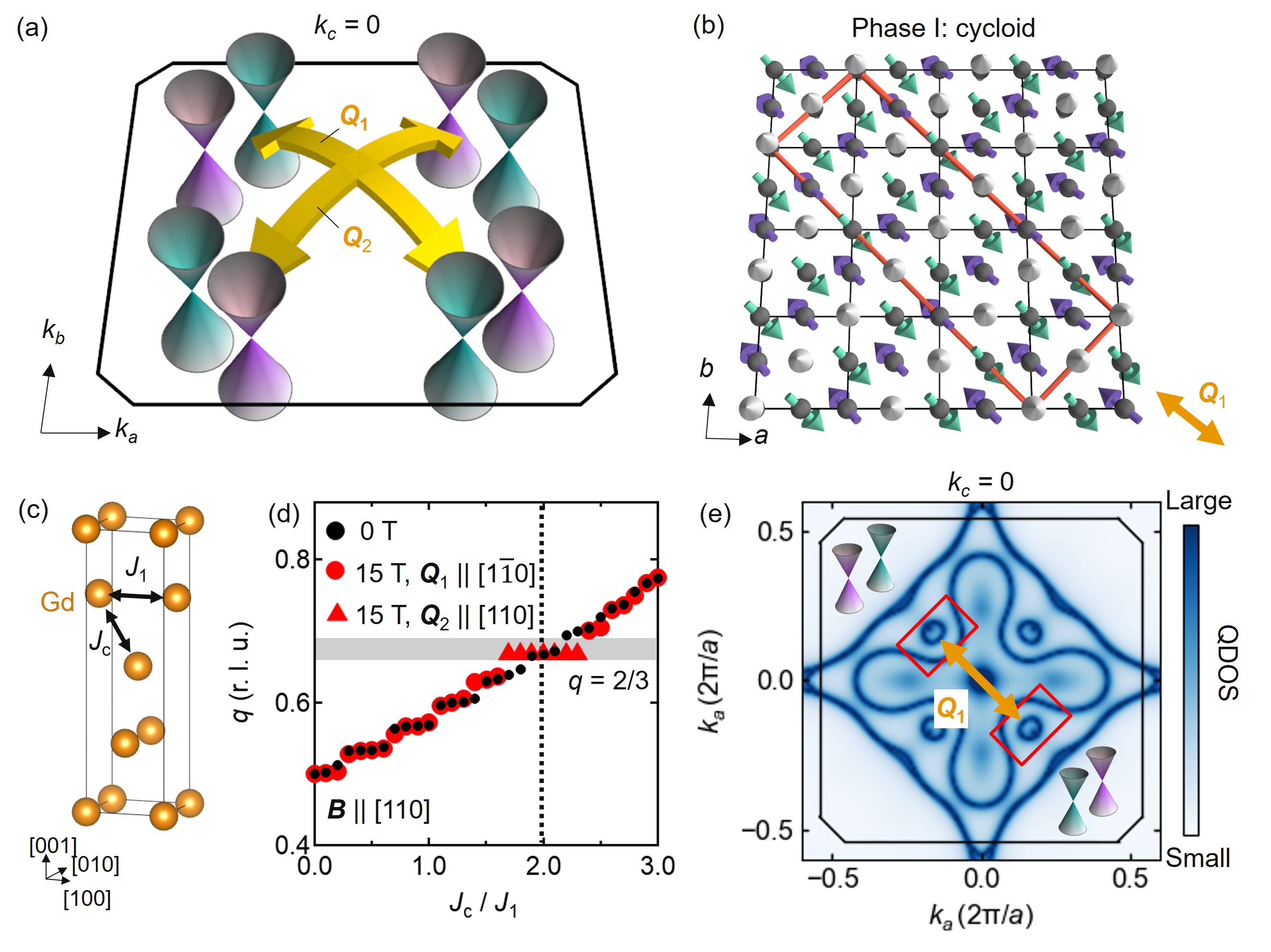}
\caption{\textbf{Undistorted helimagnetism in the Weyl semimetal GdAlSi.} (a) Illustration of magnetic modulation vectors $\bm{Q}_1$, $\bm{Q}_2$ connecting between Weyl nodes in the electronic structure of GdAlSi. The black box indicates the $k_c=0$ cut of the tetragonal Brillouin zone (BZ) in the paramagnetic state.
(b) Experimentally observed cycloidal helimagnetic structure in the ground state of GdAlSi ($B = 0$) projected onto the $(001)$ plane. Only magnetic Gd sites are shown; the black (red) boxes indicate a single unit cell in the paramagnetic (in the helimagnetic) state.
(c) Gadolinium sublattice in the $I4_1md$ tetragonal structure of GdAlSi with definitions for nearest-neighbor ($J_1$) and next-nearest neighbor interactions ($J_c$). (d) Evolution of magnetic modulation vector $\bm{Q}_1 = (q, -q, 0)$  and $\bm{Q}_2 = (q, q, 0)$ for $J_1, J_c<0$. A helimagnetic structure with $q = 2/3$ can be realized in this simple framework. 
(e) Calculated, momentum-resolved density of states (QDOS) in the $k_c=0$ plane of the tetragonal BZ. 
}

\label{Fig1}
\end{figure}

\textbf{Electronic and magnetic state of GdAlSi.}\\
GdAlSi crystallizes in the polar tetragonal $I4_1md$ structure with neither inversion center nor $\mathcal{M}_z$ mirror plane, leading to a Weyl semimetallic state that is
strongly coupled to the magnetic texture, see Fig. \ref{Fig1}a,b and Supplementary Fig. \ref{FigS_SHG}~\cite{Xu_2017_Weyl_LaAlGe,Chang_2018_RAlGe_MagWeyl,suzuki_2019_CeAlGe_AMR,gaudet_2021_NdAlSi,yao_2023_SmAlSi,li_2023_NdAlSi_ARPES,laha_2024_GdAlSi,Nag_2024_GdAlSi}. In this material class, ferromagnetism, canted ferromagnetism, topological antimerons, nearly collinear up-up-down order, and helimagnetic order have been reported from experiments on PrAlSi/Ge~\cite{Yang_2020_PrAlSiGe}, CeAlSi~\cite{Yang_2021_CeAlSi}, CeAlGe~\cite{Puphal_2020_CeAlGe}, NdAlSi/Ge~\cite{gaudet_2021_NdAlSi,Yang_2023_NdAlGe}, and SmAlSi~\cite{yao_2023_SmAlSi}, respectively. However, there is no prior confirmation of an undistorted helimagnetic order by polarization analysis, hindering discussion of the relevant exchange interactions.

Figure \ref{Fig1}a illustrates a subset of these electronic Weyl points and their approximate position close to the $k_c = 0$ plane of the tetragonal Brillouin zone in the paramagnetic state of noncentrosymmetric GdAlSi. In this article, we demonstrate the helimagnetic order in GdAlSi, which interconnects the Weyl points as indicated by yellow arrows in Fig. \ref{Fig1}a, and which has the character of a harmonic cycloid defined by  
\begin{equation}
\bm{m}(\bm{r}) = m(\bm{Q})\left[\bm{e}_{[110]}\cos\left(\bm{Q}\cdot \bm{r}\right) + \bm{e}_{c}\sin\left(\bm{Q}\cdot \bm{r}\right)\right]
\end{equation}
Here, $\bm{e}_{[110]}$ and $\bm{e}_{c}$ are unit vectors along the crystallographic $[110]$- and $c$-axes, respectively; $\bm{Q} = (q, q, 0)$ with $q=0.673(2)$ reciprocal lattice units (r.l.u.) is a lattice-incommensurate propagation vector. This cycloidal texture and the corresponding magnetic unit cell are illustrated in Fig. \ref{Fig1}b, where the staggered square lattices of gadolinium ions in GdAlSi are projected onto a single plane. The conventional unit cell of the magnetic state is about six times larger than the conventional unit cell of the paramagnetic state; primitive cells are discussed in Supplementary Fig. \ref{FigS_unit_Brillouin}. 

The helimagnetism in GdAlSi is readily understood from the viewpoint of a minimal exchange Hamiltonian with nearest-neighbor interaction $J_1<0$ and inter-layer coupling $J_c<0$, illustrated in Fig. \ref{Fig1}c. Consistent with the experiment, this natural choice of interactions (i) selects only propagation vectors of the type $\bm{Q}_{1,2}= (\pm q, \pm q, 0)$, constrained by tetragonal symmetry and (ii) reproduces the experimentally relevant range of $q$ for a reasonable $J_c/J_1$ ratio (see Fig. \ref{Fig1}d). Details of the calculation are given in Supplementary Note \ref{Esec:Spin_Hamiltonian}. 

Importantly, the $\bm{Q}$-vector observed experimentally in GdAlSi is well matched to the distance between sets of Weyl points in the electronic structure, as shown in our calculations of Fig. \ref{Fig1}e. Prior work on NdAlSi has demonstrated a modification of quantum oscillations at the Weyl nodes below the onset of magnetic order~\cite{gaudet_2021_NdAlSi}. The reciprocal effect of relativistic fermions on the magnetic structure, a strong contribution to magnetic interactions by Weyl electrons via modified Ruderman-Kittel-Kasuya-Yosida interactions, has not yet been conclusively demonstrated~\cite{Hirschberger2021, Chang2015, Hosseini2015,Araki2016}. Indeed, electronic structure calculations on $R$AlSi ($R$ = rare earth) show that states far from the Fermi energy predominantly fix the value of $\bm{Q}$~\cite{bouaziz_2024_RAlSi_mag}. Therefore, Weyl nodes and helimagnetic order both exist in GdAlSi, but helimagnetism is not predominantly caused by the Weyl electrons and the matching of $\bm{Q}$ to the separation of Weyls points in momentum space is accidental.\\

\begin{figure}[h]%
\centering
\includegraphics[width=1.0\textwidth]{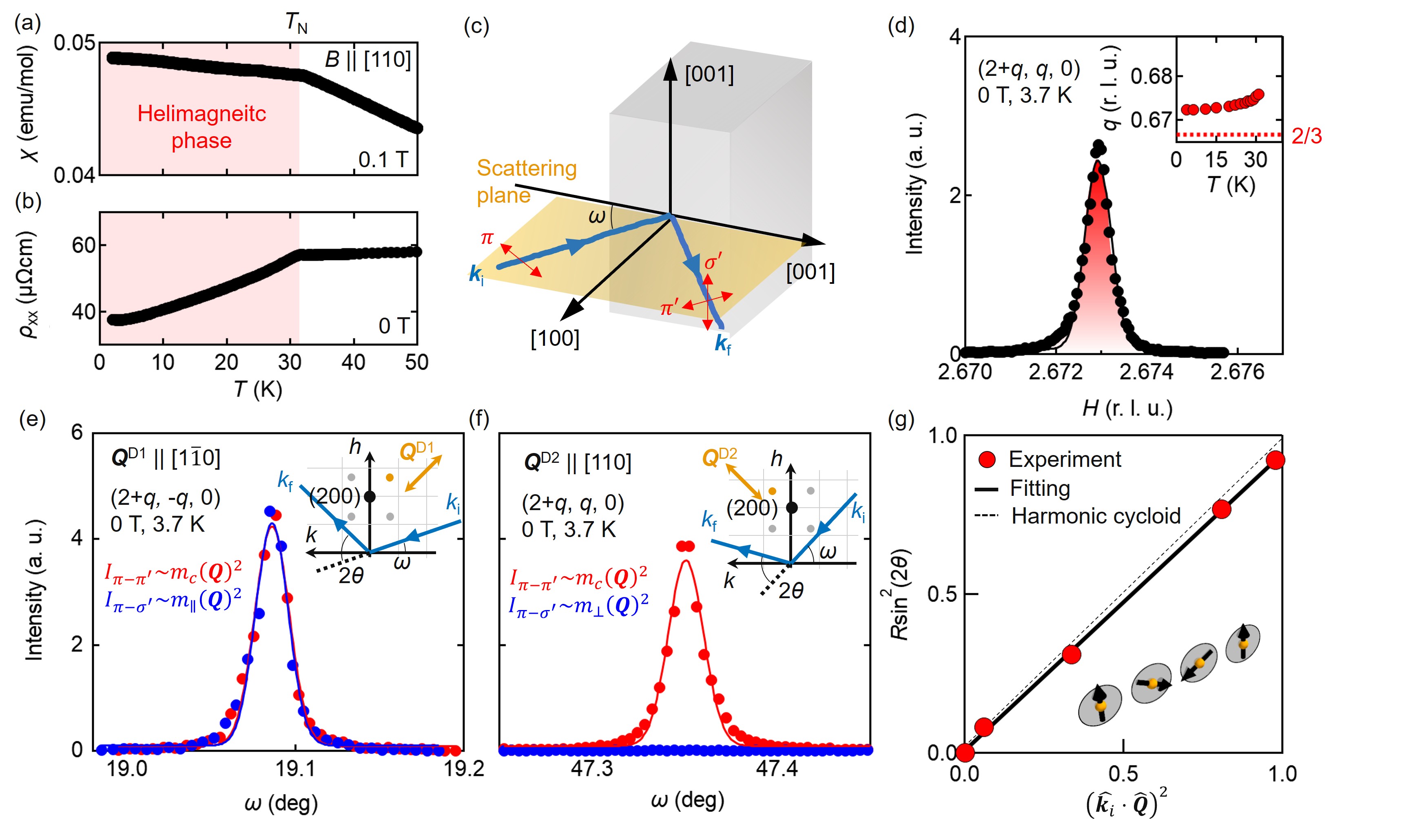}
\caption{\textbf{Undistorted helimagnetic ground state of GdAlSi from resonant elastic X-ray scattering (REXS).}
(a) Temperature ($T$-) dependent magnetic susceptibility $\chi$ in a small magnetic field applied; The N{\'e}el point $T_\mathrm{N}$ is indicated. (b) Metallic, $T$-dependent resistivity $\rho_{xx}$ with a decrease below $T_\mathrm{N}$. 
(c) Experimental configuration of resonant elastic X-ray scattering (REXS) measurements 
(Methods).
The scattering plane (yellow) is perpendicular to $[001]$. $\pi$ ($\pi'$, $\sigma'$) represent the polarization direction of the incident (scattered) X-ray beam with wave vector $\bm{k}_{\mathrm{i}}$ ($\bm{k}_{\mathrm{f}}$), respectively. 
X-rays are incident on the sample surface at an angle $\omega$ (Methods).
(d) REXS line scan profile of the $(2+q, q, 0)$ magnetic reflection without analyzer plate, where $q=2/3+\delta$ and $\delta = 0.006$; 
Gaussian fit is shaded in red. 
Inset: $T$-dependence of $q$ in $\bm{Q}_2 = (q, q, 0)$, evidencing incommensurability with the lattice. 
(e,f) Polarization analysis in REXS, assuming equal population of helimagnetic domains in zero magnetic field.
(e) Rocking scan ($\omega$) profile around $(2+q, -q, 0)$
, which is a contribution from the $\bm{Q}^{\mathrm{D1}}\parallel [1\overline{1}0]$ domain. 
$I_{\pi\text{-}\pi'}$ and $I_{\pi\text{-}\sigma'}$ reflect $m_{c}(\bm{Q})^{2}$ and $m_{\parallel}(\bm{Q})^{2}$, respectively, where $m_{c}(\bm{Q})$, $m_{\parallel}(\bm{Q})$, and $m_{\perp}(\bm{Q})$ are Fourier modes of the magnetic moment along the $[001]$ direction, along $\bm{Q}$, and along the axis perpendicular to both of them, respectively (Eq. (\ref{eq_REXS_spin})). Inset: experimental setup projected onto the $HK0$ scattering plane. 
(f) The corresponding rocking scan of $(2+q, q, 0)$ originating from the $\bm{Q}^{\mathrm{D2}}\parallel [110]$ domain. 
Here, $I_{\pi\text{-}\pi'}$ and $I_{\pi\text{-}\sigma'}$ mainly reflect $m_{c}(\bm{Q})^{2}$ and $m_{\perp}(\bm{Q})^{2}$, respectively. 
(g) Undistorted helimagnetic cycloid structure obtained by polarization analysis in REXS at various magnetic reflections. The model expectation for a harmonic cycloid is indicated by the dashed line. Inset: helimagnetic cycloidal texture.
}  
\label{Fig2}
\end{figure}

\textbf{Zero-field cycloid from resonant elastic X-ray scattering (REXS).}\\
From magnetic susceptibility measurements and the resistivity $\rho_{xx}$ in GdAlSi, we demonstrate a conducting ground state with a magnetic N{\'e}el point $T_\mathrm{N}= 32\,$K (Fig. \ref{Fig2}a,b). The magnetic susceptibility has very weak anisotropy both above and below $T_\mathrm{N}$ (see Fig. \ref{FigS_aniso}); $\rho_{xx}$ appears affected by magnetic fluctuations above and around $T_\mathrm{N}$, but fluctuations freeze out upon cooling.
We employ resonant elastic X-ray scattering (REXS) measurements 
to reveal the helimagnetic order of GdAlSi (Methods). Figure \ref{Fig2}c illustrates the experimental geometry of the sample with respect to the X-ray beam. Satellite magnetic peaks for two magnetic domains D1, D2 in vicinity of the fundamental Bragg reflection $(2, 0, 0)$ are observed with the propagation vector along the $\langle110\rangle$ directions: $\bm{Q}^{\mathrm{D1}} = (q, -q, 0)$ or $\bm{Q}^{\mathrm{D2}} = (q, q, 0)$. 
Figure \ref{Fig2}d indicates that $q$ is not perfectly matched to the lattice, i.e., incommensurate, 
in GdAlSi 
: $q = 2/3 + \delta$ with $\delta = 0.006$.

Polarization analysis of the scattered X-rays reveals that the zero-field state of GdAlSi is the only known, undistorted helimagnetic spin texture in a magnetic Weyl semimetal~\cite{gaudet_2021_NdAlSi,yao_2023_SmAlSi}. In general, the Fourier transformed magnetic moment $\bm{m}(\bm{Q})$ is separated into three mutually orthogonal components:
\begin{equation}
\bm{m(\bm{Q})} = m_{c}(\bm{Q})\bm{e}_{c} + m_{\parallel}(\bm{Q})\bm{e}_{Q} +  m_{\perp}(\bm{Q})\bm{e}_{c}\times \bm{e}_{Q},
\label{eq_REXS_spin}
\end{equation}
where $\bm{e}_{c}$ and $\bm{e}_{Q}$ are unit vectors along the $[001]$ direction of the tetragonal structure and along $\bm{Q}$, respectively. In our experiment, the incident X-rays are linearly polarized with their electric field component within the scattering plane ($\pi$-plane). The scattered X-rays can have two polarization components: parallel ($\pi'$) and perpendicular ($\sigma'$) to the scattering plane. These two polarization components are separated at the detector (Methods). 
In our geometry, the $\pi\text{-}\pi'$ scattered intensity always detects $m_{c}(\bm{Q})$; we further select two representative magnetic reflections for which the intensity of the $\pi\text{-}\sigma'$ channel corresponds to $m_{\parallel}(\bm{Q})$ and $m_{\perp}(\bm{Q})$ in Eq. (\ref{eq_REXS_spin}), respectively.
Figure \ref{Fig2}e,f suggests that GdAlSi has $m_{c}(\bm{Q})$ and $m_{\parallel}(\bm{Q})$ of comparable magnitude, but not $m_{\perp}(\bm{Q})$: this corresponds to a harmonic helimagnetic cycloidal structure at zero magnetic field. This incommensurate magnetic cycloid has face-centered orthorhombic symmetry (magnetic space group $Fdd2$)\footnote{When approximated as a commensurate structure with $q = 2/3$, the magnetic space group is $Fd^\prime d^\prime2$.}. \\

\begin{figure}[h]%
\centering
\includegraphics[width=1.0\textwidth]{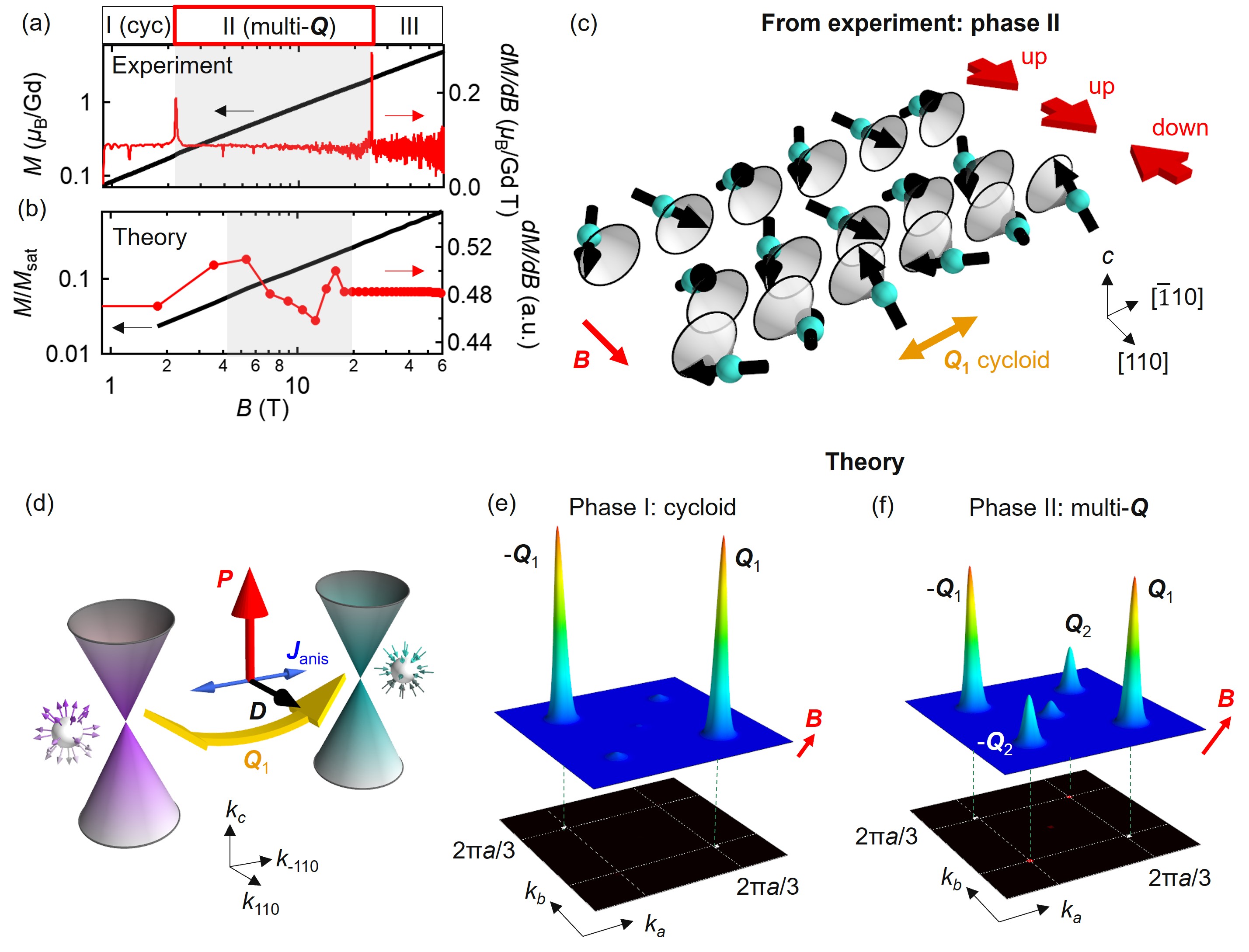}
\caption{\textbf{Field-induced multi-$Q$ spin textures and magnetic model calculations for the Weyl semimetal GdAlSi}. (a,b) Magnetization isotherms at $T = 4.2\,$K from experiment and model calculations. The magnetic field $\bm{B}$ is along the $[110]$ axis. Two $B$-induced transitions are visible as sharp maxima in $dM/dB$. (c) Multi-$\bm{Q}$ magnetic texture observed in phase II of GdAlSi. The structure is composed of an incommensurate, harmonic cycloid along $\bm{Q}_1$ and a commensurate, collinear up-up-down texture with moments parallel to $\bm{B}$. See Supplementary Note \ref{SNote_HF_spin} for detailed experiments on this structure. (d) Illustration of two Weyl cones of opposite charge and the magnetic interactions allowed for the polar (polarization $\bm{P}\parallel [001]$) crystal structure of this material: The Dzyaloshinskii-Moriya (DM) interaction $\bm{D}\perp \bm{Q}$ and the anisotropic exchange interaction $J_\mathrm{anis}$, which favor cycloidal and proper screw habits, respectively. (e,f) Magnetic scattering intensity $I(\bm{Q})$ in the $k_c = 0$ plane calculated from Eq. (\ref{eq_Kondo}) in the $B = 0$ ground state and in the field-induced state; the undistorted cycloid and the magnetic texture of panel (c) are well reproduced. }
\label{Fig3}
\end{figure}

\textbf{Multi-$Q$ spin textures of GdAlSi induced by a magnetic field.}\\
We apply a magnetic field to the cycloidal helimagnetic ground state of GdAlSi, $\bm{B} \parallel [110]$, and observe two field-induced transitions in Fig. \ref{Fig3}a. Strictly speaking, the critical field for the transition into the field-aligned ferromagnetic state, $B_c\approx 75\,$T, is beyond our experimental range. We estimate $B_c$ by linear extrapolation of $M(B)$ to the saturation magnetization of Gd$^{3+}$, $7\,\mathrm{\mu_B}$. Using REXS with polarization analysis in an applied field of $7\,$T along the $[110]$ direction, we determine the magnetic structure of phase II (Supplementary Fig. \ref{FigS_HF_spin}). As summarized by the illustration in Fig. \ref{Fig3}c, this is a superposition of two magnetic waves, termed a multi-$\bm{Q}$ state: the dominant component is an incommensurate magnetic cycloid along $\bm{Q}_1 = (q, -q, 0)$ with $q \approx 0.673(2)$, whose spin components are in the $[1\bar{1}0]$-$c$ plane, perpendicular to $\bm{B}$. This is paired with a subdominant up-up-down pattern along the commensurate $\bm{Q}_2 = (2/3, 2/3, 0)$ for the spin component parallel to $\bm{B}$, $\bm{m}_{[110]}(\bm{Q})$. The combination of incommensurate and commensurate modulations can be depicted as a staggered cone structure, as in Fig. \ref{Fig3}c.\\

\textbf{Spin model calculations.}\\
GdAlSi's simple helimagnetic cycloidal order in zero magnetic field can be explained from a simple model of frustrated isotropic exchange interactions, as in Fig. \ref{Fig1}c,d, when adding Dzyaloshinskii-Moriya interactions (DMI) that are allowed by the symmetry of the polar structure~\cite{bouaziz_2024_RAlSi_mag}. 
However, to reproduce the helimagnetic multi-$\bm{Q}$ textures under a magnetic field, we require both DMI and anisotropic exchange interactions. This is discussed in the following. Figure \ref{Fig3}d illustrates the relative alignment of the ordering vector $\bm{Q}$, the polarization $\bm{P}$ imposed by the $I4_1md$ crystal structure, the Dzyaloshinskii-Moriya (DM) vector $\bm{D}$, and the anisotropic exchange $J_\mathrm{anis}$. Consider the example of $\bm{Q}_2\parallel [110]$: Here, the relevant $\bm{D}$ couples the magnetization components $m_{[110]}(\bm{Q}_2)$ and $m_c(\bm{Q}_2)$, whereas $J_\mathrm{anis}$ couples $m_{[110]}(\bm{Q}_2)$ and $m_{[1\bar{1}0]}(\bm{Q}_2)$. 

We use an effective Fourier-space Hamiltonian on a square lattice, which includes both DMI and anisotropic exchange interactions in the matrix $\Gamma^{\alpha \beta}_{\bm{Q}_{\nu}}$~\cite{hayami_2024_Qmodel_review}, viz.
\begin{equation}
\mathcal{H} = -2
\sum_{\nu,\alpha,\beta} \Gamma^{\alpha \beta}_{\bm{Q}_{\nu}} m_{\alpha}(\bm{Q}_{\nu})m_{\beta}(\bm{Q}_{\nu}) - \sum_{i} \bm{B} \cdot \bm{m}(\bm{r}_{i}),
\label{eq_Kondo}
\end{equation}
where $m_{\alpha}(\bm{Q}_{\nu})$ with $\alpha=x,y,z$ and $\nu=1$--$8$ are Fourier components of the localized magnetic moment $\bm{m}(\bm{r}_{i})$; $\nu=5-8$ are commensurate with the lattice, while $\nu = 1-4$ are incommensurate (Methods). The second term represents the Zeeman energy with $\bm{B}=(B, B, 0)/\sqrt{2}$. 

In Fig. \ref{Fig3}e,f we show the calculated magnetic scattering intensities, i.e., the square of the magnetic structure factor, for phases I and II. The model reproduces the cycloidal helimagnetic ground state, with two incommensurate Fourier components $\pm \bm{Q}_1$. Moreover, the model also features the field-induced phase II, with two strong and incommensurate reflections at $\pm \bm{Q}_1$ of cycloidal spin habit, combined with two weaker commensurate reflections at $\pm \bm{Q}_2$ with collinear spin habit (spins parallel to $\bm{B}$). In other words, the calculation is well consistent with the experimental observations in Fig. \ref{Fig3}c and Supplementary Fig. \ref{FigS_HF_spin}. This good alignment between theoretical and experimental results, as a function of magnetic field, supports the accuracy of our magnetic structure analysis in Figs. \ref{Fig2}, \ref{Fig3} and in Supplementary Fig. \ref{FigS_HF_spin}. It also indicates that the Hamiltonian of Eq. (\ref{eq_Kondo}) is suitable to describe magnetic order in GdAlSi. In Supplementary Note \ref{Esec:Spin_Hamiltonian}, we further truncated the Fourier-space model of Eq. (\ref{eq_Kondo}) after a few near-neighbours and obtained an intuitive real-space Hamiltonian , to discuss the energetics of the helimagnetic ground state. \\

\textbf{Discussion and Conclusion.}\\ 
Our Hamiltonian model focuses on Fourier components with characteristic propagation vectors $\bm{Q}_\nu$, which are well matched to the Fermi surface of GdAlSi in Fig. \ref{Fig1}e. Here, pairs of Weyl nodes facing each other along the $[110]$ direction are connected by $\bm{Q}$, c.f. Fig. \ref{Fig1}a, and we expect a strong response of the electron gas when the helimagnetic $\bm{Q}$-vector changes its direction. Indeed, controlling these $\bm{Q}$-domains in magnetic Weyl semimetals, by magnetic field or current, provides a valley-selective way to open a (partial) charge gap around the Weyl cones, if they have an energy-like causal connection in the folded Brilloin zone~\cite{Chiu2023}. 

Our study also emphasizes the role of symmetry-allowed Dzyaloshinskii-Moriya (DM) interactions, which favor the cycloidal spin habit, combined with anisotropic exchange, which stabilizes the multi-$\bm{Q}$ order in a magnetic field. Note that the spiral-type DM interaction, which has been proposed for magnetic Weyl systems, is strictly symmetry-forbidden by the polar crystal structure of $R$AlSi, $R=\,$rare earth~\cite{Chang2015, Araki2016}. However, spiral states can be stabilized by the presently observed anisotropic exchange interactions and may play a role not only in NdAlSi~\cite{gaudet_2021_NdAlSi} but also in the wider class of inversion breaking Weyl semimetals.



\newpage
\begin{center}
\Large{References}
\end{center}
\bibliography{sn-bibliography}


\newpage

\newpage
\begin{center}
\Large{Methods}
\end{center}

\textbf{Crystal growth and characterization.}\\
Single crystals of GdAlSi were synthesized by the Al flux method. High purity chunks of Gd, Al, and Si were loaded into alumina crucibles with molar ratio Gd:Al:Si = $1:20:2$, sealed in an evacuated quartz tube, heated to $1175\,^\circ$C, held at this temperature for $12$ h, cooled to $700\,^\circ$C at $0.7\,^\circ$C/h, kept at $700\,^\circ$C for $12$ h, and centrifuged to remove residual Al flux. The samples were characterized by powder X-ray diffraction (XRD) of crushed single crystals and by Laue X-ray diffraction of single crystals. \\

\textbf{Magnetization and electrical transport measurements.} \\
Magnetization was measured with a Magnetic Property Measurement System (MPMS, Quantum Design) and with the vibrating sample magnetometer (VSM) option of a Physical Property Measurement System (PPMS, Quantum Design). 
The $M$-$H$ curve up to $60\,\mathrm{T}$ in Fig. \ref{Fig3}a was measured in a pulsed magnetic field using an induction method with coaxial pickup coils.
Electric transport measurements with rectangular shaped samples were performed using the PPMS. A standard four-probe method was applied to enable precise transport measurements irrespective of contact resistances.
\\

\textbf{Resonant elastic X-ray scattering (REXS).}\\
REXS measurements were performed at beamline BL19LXU of SPring-8 and BL-3A of Photon Factory, KEK. The photon energy of the incident X-rays was tuned to the Gd $L_2$ absorption edge ($\sim7.936\,$keV). For the zero field measurements at SPring-8, a single crystal of GdAlSi with a polished (100) surface was set in a cryostat so that the scattering plane was $(H, K, 0)$. For the finite field measurements at KEK, we set the sample in a vertical-field superconducting magnet with the $(H, H, L)$ scattering plane, so that the magnetic field was applied along the $[110]$ direction. At both SPring-8 and KEK, the incident X-rays had linearly polarized electric field in the scattering plane ($\pi$-polarization). The (006) reflection of a pyrolytic graphite (PG) plate with $2\theta = 88\,^\circ$ was used to detect the polarization component of the scattered X-rays in the scattering plane ($\pi^{\prime}$) and out of the scattering plane ($\sigma^{\prime}$), by rotating the PG plate around the X-ray beam. In general, when the magnetic structure is described by the Fourier component of the magnetic moment $\bm{m}(\bm{Q})$, the resonantly enhanced intensity is written as
\begin{equation}
I\propto\lvert(\bm{\varepsilon}_{\mathrm{i}}\times\bm{\varepsilon}_{\mathrm{f}})\cdot\bm{m}(\bm{Q})\rvert^{2},
\end{equation}
where $\bm{\varepsilon}_{\mathrm{i}}$ and $\bm{\varepsilon}_{\mathrm{f}}$ represent the polarization vectors of the incident and scattered X-rays~\cite{REXS_textbook}, respectively. In our experiments, the intensity $I_{\pi\text{-}\pi'}$ for the $\pi\text{-}\pi'$ channel is proportional to $\left|m_{\mathrm{out}}(\bm{Q})\right|^{2}$ and $I_{\pi\text{-}\sigma'}$ for the $\pi\text{-}\sigma'$ channel is proportional to $(\bm{m}_{\mathrm{in}}(\bm{Q})\cdot\bm{k}_{\mathrm{i}})^{2}$, where $\bm{m}_{\mathrm{in}}(\bm{Q})$ and $m_{\mathrm{out}}(\bm{Q})$ are magnetic moments in the scattering plane and perpendicular to it. $m_{\mathrm{out}}(\bm{Q})$ corresponds to $m_{c}(\bm{Q})$, and $\bm{m}_{\mathrm{in}}(\bm{Q})$ are separated into $m_{\parallel}(\bm{Q})$ and $m_{\perp}(\bm{Q})$ in the Main Text.\\

\textbf{Fourier-space model calculations.}
\\
The data in Fig. \ref{Fig3}e,f are obtained by simulated annealing for an effective spin Hamiltonian on a two-dimensional square lattice,
\begin{equation}
\mathcal{H} = -2
\sum_{\nu,\alpha,\beta} \Gamma^{\alpha \beta}_{\bm{Q}_{\nu}} m_{\alpha}(\bm{Q}_{\nu})m_{\beta}(\bm{Q}_{\nu}) - \sum_{i} \bm{B} \cdot \bm{m}(\bm{r}_{i}),
\label{eq_Kondo2}
\end{equation}
where $m_{\alpha}(\bm{Q}_{\nu})$ with $\alpha=x,y,z$ and $\nu=1$--$8$ are Fourier components of the localized magnetic moment $\bm{m}(\bm{r}_{i})$ and $\Gamma^{\alpha \beta}_{\bm{Q}_{\nu}}$ is a matrix containing the DMI and anisotropic exchange interaction, as is explained in the following. The second term represents the Zeeman energy with $\bm{B}=(B, B, 0)/\sqrt{2}$. $N=60^2$ is the total number of spins and the amplitude of each spin is fixed to unity. 
To obtain the low-energy spin configuration, the temperature is gradually reduced from $T=1$ to $T=0.001$ with a ratio $T_{n+1}=\alpha T_n$, where $T_n$ is the $n$th-step temperature and $\alpha$ is set between 0.999990 and 0.999999. 
At each temperature, a local spin update is performed in real space based on the standard Metropolis algorithm. 
At the final temperature, $10^5$--$10^6$ Monte Carlo sweeps are performed for the measurement.

The explicit propagation vectors in Eq. (\ref{eq_Kondo2}) are: $\bm{Q}_1=(q, q, 0)$, $\bm{Q}_2=(-q, q, 0)$, $\bm{Q}_3=(q', q', 0)$, $\bm{Q}_4=(-q',q',0)$, $\bm{Q}_5=2\bm{Q}_1+\bm{Q}_4$, $\bm{Q}_6=2\bm{Q}_1-\bm{Q}_4$, $\bm{Q}_7=2\bm{Q}_2+\bm{Q}_3$, and $\bm{Q}_8=2\bm{Q}_2-\bm{Q}_3$ with $q=0.35$ and $q'=1/3$. Here, $\bm{Q}_1$ and $\bm{Q}_2$ correspond to the incommensurate wave vectors, $\bm{Q}_3$ and $\bm{Q}_4$ correspond to the commensurate wave vectors, and $\bm{Q}_5$-$\bm{Q}_8$ correspond to higher harmonic wave vectors that assist the stabilization of the field-induced distorted helimagnetic state consisting of several spin density waves. To map the three-dimensional spin structures onto the two-dimensional square lattice, $q = 0.35$ and $q' = 1/3$ are incorporated, instead of the experimentally observed $q = 2/3 + \delta$ ($\delta = 0$ and $0.006$). 

The model parameters of the exchange matrix in Eq. (\ref{eq_Kondo}) are set to satisfy the effective $4$-fold rotational symmetry of the crystal structure of GdAlSi as follows~\cite{yambe_2022_Qsquare}: The isotropic exchange interaction is given by $\Gamma^{xx}_1=\Gamma^{yy}_1=\Gamma^{zz}_1=\Gamma^{xx}_2=\Gamma^{yy}_2=\Gamma^{zz}_2 \equiv J$, $\Gamma^{xx}_3=\Gamma^{yy}_3=\Gamma^{zz}_3=\Gamma^{xx}_4=\Gamma^{yy}_4=\Gamma^{zz}_4 \equiv \alpha_1 J$, and $\Gamma^{xx}_5=\Gamma^{yy}_5=\Gamma^{zz}_5=\Gamma^{xx}_6=\Gamma^{yy}_6=\Gamma^{zz}_6=\Gamma^{xx}_7=\Gamma^{yy}_7=\Gamma^{zz}_7=\Gamma^{xx}_8=\Gamma^{yy}_8=\Gamma^{zz}_8 \equiv \alpha_2 J$, the DMI is given by 
$\Gamma^{zx}_1=-\Gamma^{xz}_1=\Gamma^{zy}_1=-\Gamma^{yz}_1=-\Gamma^{zx}_2=\Gamma^{xz}_2=\Gamma^{zy}_2=-\Gamma^{yz}_2 \equiv D $ and 
$\Gamma^{zx}_3=-\Gamma^{xz}_3=\Gamma^{zy}_3=-\Gamma^{yz}_3=-\Gamma^{zx}_4=\Gamma^{xz}_4=\Gamma^{zy}_4=-\Gamma^{yz}_4 \equiv \alpha_1 D $, and the symmetric anisotropic interaction is given by 
$\Gamma^{xy}_1=\Gamma^{yx}_1=-\Gamma^{xy}_2=-\Gamma^{yx}_2 \equiv E$ and $\Gamma^{xy}_3=\Gamma^{yx}_3=-\Gamma^{xy}_4=-\Gamma^{yx}_4 \equiv \alpha_1 E'$ with $J=1$, $\alpha_1=0.98$, $\alpha_2=0.95$, $D=0.01$, $E=0.045$, and $E'=0.05$. 
\\

\textbf{Electronic structure calculations.}\\
First-principles calculations of the electronic structure were carried out using the all-electron full-potential Korringa-Kohn-Rostoker (KKR) Green's function method~\cite{russmann2022juktkr}. Relativistic spin-orbit coupling was included self-consistently, and the Gd atoms were treated with the generalized gradient approximation (GGA) exchange-correlation functional~\cite{perdew_1996_GGA}. The angular momentum cutoff for the Green’s function orbital expansion was set to \(l_{\text{max}} = 3\), and the complex energy contour was defined using $51$ integration points. To reduce computational cost, the calculations are carried out based on the primitive crystallographic unit cell, not the conventional unit cell (Supplementary Fig. \ref{FigS_unit_Brillouin}).
Self-consistent calculations for the primitive cell were performed with a $\bm{k}$-mesh of $30 \times 30 \times 30$. 
For Fig. \ref{Fig1}e, Fermi surface cuts were computed from the quasiparticle density of states (QDOS) in the fully polarized ferromagnetic state~\cite{ebert_2011_QDOS}.\\

\textbf{Second harmonic generation (SHG).}\\
SHG measurements are performed to show inversion-symmetry breaking of bulk GdAlSi in the paramagnetic phase at room temperature. 
As a light source, we used a Ti:sapphire oscillator (MaiTai HP, Spectra Physics) 
with a pulse duration of $100\,\mathrm{fs}$, repetition rate of $80\,\mathrm{MHz}$,
and center wavelength of $800\,\mathrm{nm}$.
The laser pulse is focused onto the $c$-plane surface of GdAlSi in $45$-degree oblique incidence and the reflected SHG is detected by a spectrometer equipped with a liquid-nitrogen-cooled charge coupled device. Figure \ref{FigS_SHG}a,b respectively show the incident light-polarization dependence of 
horizontally polarized ($P_\mathrm{out}$) and vertically polarized ($S_\mathrm{out}$) SHG intensity at room temperature; 
the output polarizer is fixed with polarization parallel to the $[101]$-axis ($P_\mathrm{out}$) and $[010]$ axis ($S_\mathrm{out}$).
We observed a clear polarization anisotropy of SHG in both cases, in accord with the inversion breaking crystal symmetry. 
In the paramagnetic phase with point group $4mm$, the horizontally polarized ($P_\mathrm{out}$) and vertically polarized ($S_\mathrm{out}$) SHG intensities, $I_P(\phi)$ and $I_S(\phi)$,  are given as,
\begin{equation}
    I_P (\phi) \propto \frac{1}{8} 
    \left[ 
        \left( 2a \chi_{xxz} + b \chi_{zxx} + c \chi_{zzz} \right) \cos^2 \phi 
        + 2d \chi_{zyy} \sin^2 \phi 
    \right]^2,
\end{equation}
\begin{equation}
    I_S (\phi) \propto 2 
    \left[ 
        e \chi_{yyz} \cos \phi \sin \phi 
    \right]^2,
\end{equation}
where $\chi_{ijk}(2\omega)$ and $\phi$ represent a second-order nonlinear optical susceptibility and the polarization angle for the fundamental light, respectively; $a$, $b$, $c$, $d$, and $e$ denote prefactors including the Fresnel coefficient, beam conditions of the incident light, and so on. As shown in Supplementary Fig. \ref{FigS_SHG}b,c, the model well reproduces the observed anisotropy of SHG polarization, which highlights inversion breaking of the crystal.\\

\begin{center}
\Large{Acknowledgments}
\end{center}
We acknowledge fruitful discussions with Max T. Birch. Support is acknowledged from the Japan Society for the Promotion of Science (JSPS) under Grant Nos. JP22H04463, JP23H05431, JP21K13873, JP22F22742, JP22K20348, JP23K13057, JP24H01607, JP24H01604, JP21H01037, JP23H04869, and JP23K13068. The work was partially supported by the Japan Science and Technology Agency via JST CREST Grant Numbers JPMJCR1874 and JPMJCR20T1 (Japan), as well as JST FOREST (JPMJFR2238, JPMJFR2366, and JPMJFR212X). We are grateful for support by the Murata Science Foundation, Yamada Science Foundation, Hattori Hokokai Foundation, Mazda Foundation, Casio Science Promotion Foundation, Inamori Foundation, Kenjiro Takayanagi Foundation, Toray Science Foundation, the Marubun Exchange Grant, the Foundation for Promotion of Material Science and Technology of Japan (MST Foundation), the Yashima Environment Technology Foundation, ENEOS Toenegeneral Research/Development Encouragement \& Scholarship Foundation, and Yazaki Memorial Foundation for Science and Technology. This work was supported by the RIKEN TRIP initiative (RIKEN Quantum, Advanced General Intelligence for Science Program, Many-Body Electron Systems). J. B. was supported by the Alexander von Humboldt Foundation through the Feodor Lynen Research Fellowship.
Resonant X-ray scattering at SPring-8 was carried out under proposal number 20220083. Resonant X-ray scattering at Photon Factory (KEK) was carried out under proposal numbers 2022G551 and 2023G611. \\

\begin{center}
\Large{Data and code availability}
\end{center}
The raw data and code used to produce the figures and conclusions presented in this study have been deposited, with detailed comments, on the \textit{Publication Data Repository System} of RIKEN Center for Emergent Matter Science (Wako, Japan). These data are available from the corresponding authors upon reasonable request. \\

\begin{center}
\Large{Acknowledgements}
\end{center}
M.H. and Y.To. conceived the project. R.N., R.Y., A.K., and Y.Tag. grew and characterized the single crystals. R.N. and  R.Y. performed all magnetic, electric measurements. R.N., R.Y., H.O., Y.Tan., and M.H. performed resonant X-ray scattering experiments at beamline BL19LXU of SPring-8. R.N., R.Y., M.G., H.S., H.N., and M.H. performed resonant X-ray scattering experiments at beamline BL-3A of Photon Factory. REXS experiments were designed and REXS data were analyzed under the guidance of T.-h. A. M.C. and J.M performed Monte Carlo calculations. K. S., Y. O., and Y. T. performed SHG measurements. S.H. performed simulated annealing. J.B. and R.A. performed first-principles calculations. R.N., R.Y., and M.H. wrote the manuscript with help of J.B., J.M., and S.H; all authors discussed the results and commented on the manuscript.\\

\begin{center}
\Large{Competing interests}
\end{center}
The authors declare no competing interests.


\clearpage
\begin{center}
\Large{Main Text Figures}
\end{center}

\clearpage

\clearpage





\clearpage
\begin{center}
\Large{Supplementary Information}
\end{center}





\stepcounter{hoge}
\renewcommand\thefigure{S\arabic{figure}} 
\renewcommand\thetable{S\arabic{table}} 
\renewcommand\thesection{S\arabic{section}} 
\renewcommand\theequation{S\arabic{equation}} 

\makeatletter
\renewcommand\@bibitem[1]{\item\if@filesw \immediate\write\@auxout
    {\string\bibcite{#1}{A\the\value{\@listctr}}}\fi\ignorespaces}
\def\@biblabel#1{S[#1]}
\makeatother

\begin{figure}[h]%
\centering
\includegraphics[width=0.9\textwidth]{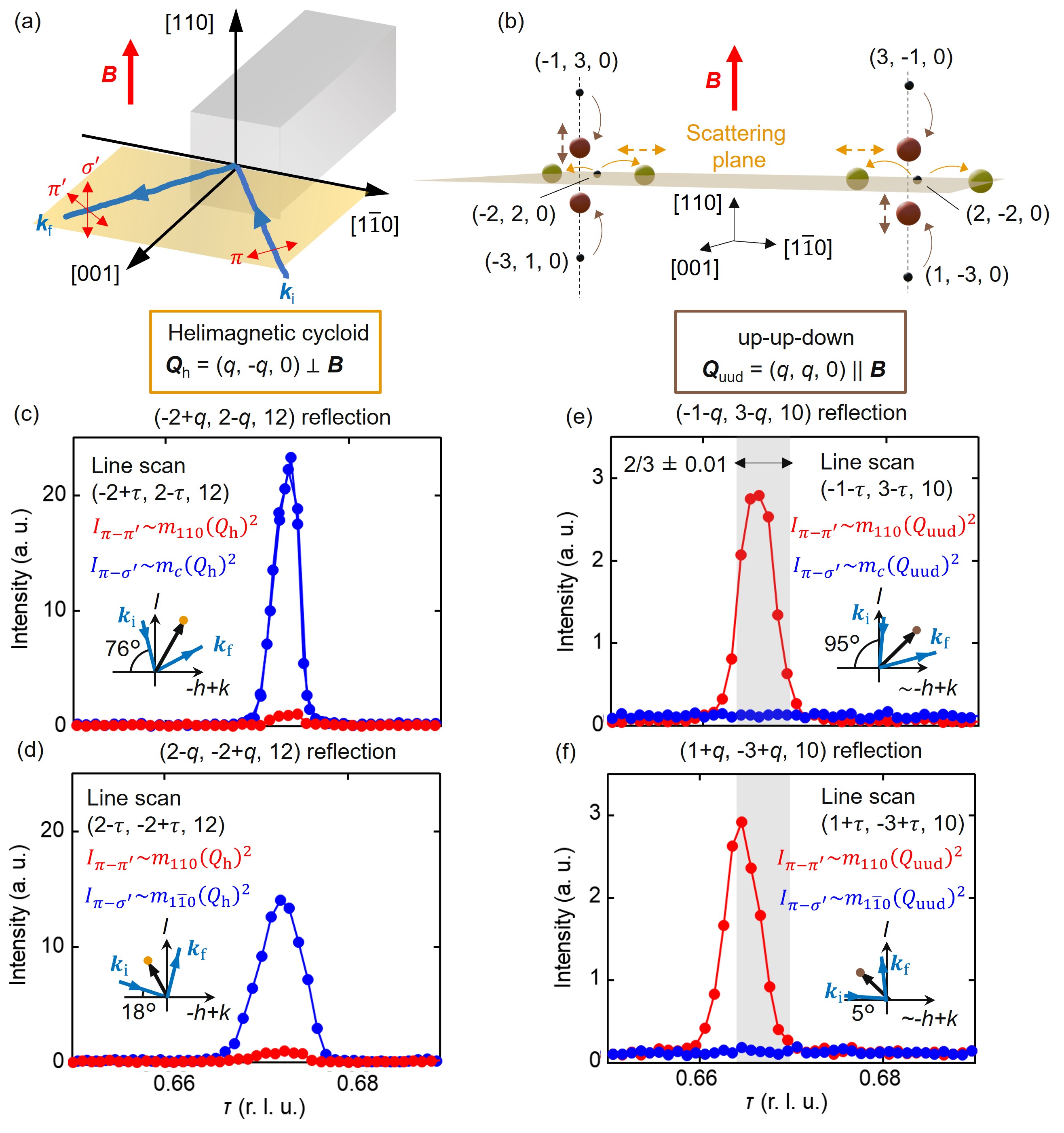}
\caption{\textbf{Magnetic structure of GdAlSi in $\bm{B}$ = 7 T external magnetic field along the [110] axis, from resonant elastic X-ray scattering (REXS).} (a) The multi-$\bm{Q}$ helimagnetic structure realized at high magnetic field is a superposition of a dominant propagation vector $\bm{Q}_\mathrm{h}$ with helimagnetic cycloidal habit, and a subdominant $\bm{Q}_\mathrm{uud}$ with up-up-down spin habit. (b) 2D contour plot of the distorted helimagnetic structure from spin model calculations. (Caption continued on following page.) 
}
\label{FigS_HF_spin}
\end{figure}

\addtocounter{figure}{-1}

\clearpage
\begin{figure}[h]%
\centering
\caption{(Continued from previous page.) (c) Experimental configuration in REXS measurements with magnetic field $\bm{B}\parallel[110]$ at BL-3A of Photon Factory, KEK (Methods). The scattering plane is perpendicular to $[110]$. $\pi$ ($\pi'$, $\sigma'$) represent the polarization direction of the incident (scattered) X-ray beam, labelled as $\bm{k}_{\mathrm{i}}$ ($\bm{k}_{\mathrm{f}}$). (d) Schematic illustration of a plane in reciprocal space perpendicular to the $[001]$ axis. Orange (brown) points indicate the magnetic reflections perpendicular (parallel) to $\bm{B}$. Black arrows indicate the directions of line scans in reciprocal space. 
(e,f) Line profiles of $(-2+\tau, 2-\tau, 12)$ and $(2-\tau, -2+\tau, 12)$ scans for $0.65 < \tau < 0.69$ around the helimagnetic cycloidal reflections ($\bm{Q}_{\mathrm{h}}\perp\bm{B}$). 
Two different reflections, $(-2+q, 2-q, 12)$ and $(2-q, -2+q, 12)$, are chosen so that $I_{\pi\text{-}\sigma'}$ from the $\pi\text{-}\sigma'$ channel mostly represents $m_{c}(\bm{Q}_{\mathrm{h}})^{2}$ and $m_{1\overline{1}0}(\bm{Q}_{\mathrm{h}})^{2}$, respectively, where $m_{c}(\bm{Q}_{\mathrm{h}})$, $m_{110}(\bm{Q}_{\mathrm{h}})$, $m_{1\overline{1}0}(\bm{Q}_{\mathrm{h}})$ represent Fourier transformed magnetic moment along the $[001]$ direction, along the $[110]$ direction, and along the $[1\overline{1}0]$ direction, respectively. $I_{\pi\text{-}\pi'}$ from the $\pi\text{-}\pi'$ channel represents $m_{110}(\bm{Q}_{\mathrm{h}})^{2}$. Sharp peaks in the $\pi\text{-}\sigma'$ channel from both reflections illustrate comparable $m_{c}(\bm{Q}_{\mathrm{h}})$ and $m_{1\overline{1}0}(\bm{Q}_{\mathrm{h}})$: helimagnetic cycloidal structure.
The inset shows the X-ray beam path projected onto the scattering plane. (g,h) Line profiles for $(-1+\tau, 3-\tau, 10)$ and $(1+\tau, -3+\tau, 10)$ scans for $0.65 < \tau < 0.69$ around the up-up-down component of the texture ($\bm{Q}_{\mathrm{uud}}\parallel\bm{B}$). 
Two different reflections, $(-1-q, 3-q, 10)$ and $(1+q, -3+q, 10)$ are chosen so that $I_{\pi\text{-}\sigma'}$ mostly represents $m_{c}(\bm{Q}_{\mathrm{uud}})^{2}$ and $m_{1\overline{1}0}(\bm{Q}_{\mathrm{uud}})^{2}$, respectively. $I_{\pi\text{-}\pi'}$ represents $m_{110}(\bm{Q}_{\mathrm{uud}})^{2}$.
Grey shading indicates the commensurate value ($q = 2/3$); r.l.u. are reciprocal lattice units. Since only the intensity from the $\pi\text{-}\pi'$ channel is present, there is only the Fourier component with moment along the [110] direction: up-up-down structure.
}
\end{figure}


\begin{figure}[h]%
\centering
\includegraphics[width=0.9\textwidth]{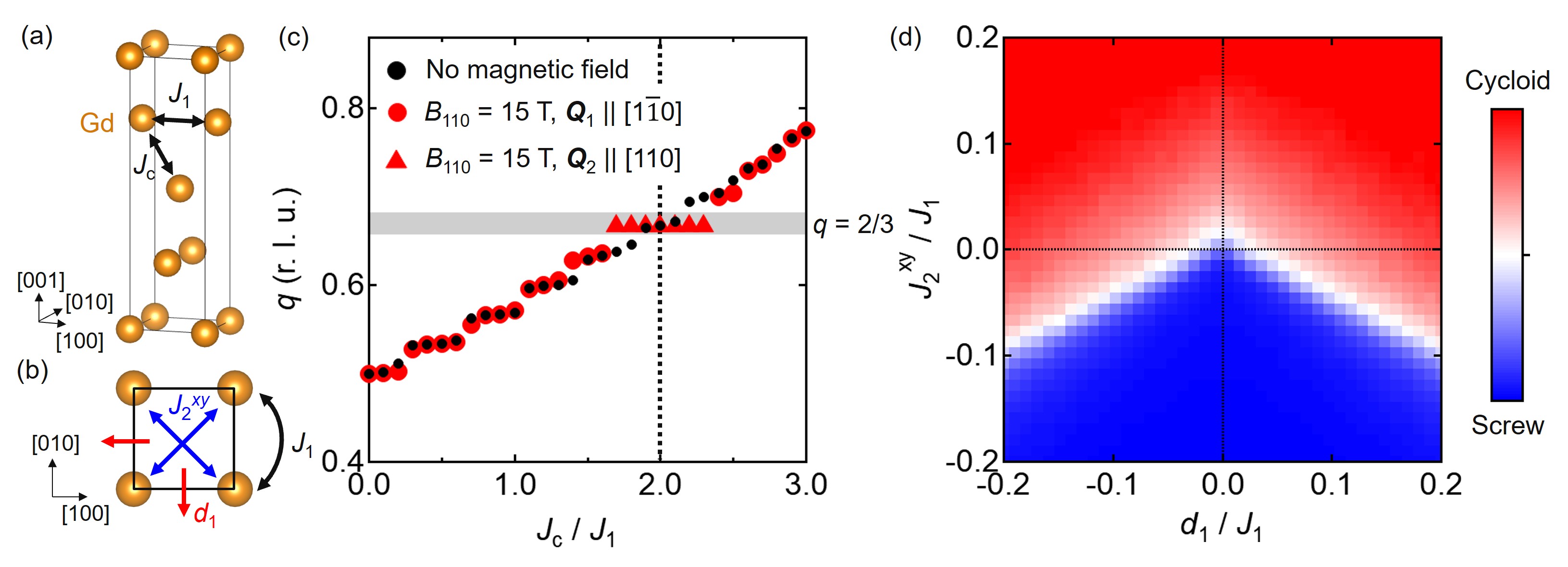}
\caption{
\textbf{Helimagnetism of GdAlSi from spin Hamiltonian in real space.} 
(a) Conventional unit cell of GdAlSi, showing only Gd atoms.
In-plane nearest neighbour interactions ($J_1$, here $J_1=1\,\mathrm{meV}$) and inter-layer nearest neighbour interactions ($J_c$) as in Monte Carlo (MC) simulations of the magnetic ground state. 
(b) Projection of a single layer of the structure in panel (a) onto the tetragonal basal plane. 
Dzyaloshinskii-Moriya interactions (DMI) for the in-plane nearest neighbour bonds ($d_1$) and anisotropic exchange interactions for the in-plane next-nearest neighbour bonds ($J_{2}^{xy}$) are depicted (Supplementary Note \ref{Esec:Spin_Hamiltonian}).  
(c) Evolution of the propagation vector $\bm{Q}_\mathrm{1} = (q, -q, 0)$ or $\bm{Q}_\mathrm{2} = (q, q, 0)$ as a function of the ratio $J_c / J_1$, from MC simulations with $d_1=J_{2}^{xy}=0$ at $T=1\,\mathrm{K}$. 
Red (black) symbols are calculated with (without) magnetic field $B_{110}=15\,\mathrm{T}$ along the $[110]$ direction. 
Red circles and triangles represent $\bm{Q}_{1}$ and $\bm{Q}_{2}$, respectively.
For $B=0\,\mathrm{T}$ (black circles), this distinction is irrelevant as the directions are equivalent.
The characteristic $q = 2/3$ (highlighted in gray) is stabilized in a region around $J_c / J_1 = 2$ at finite field $B_{110}$. 
(d) With only isotropic Heisenberg interactions in zero magnetic field, helimagnetic cycloidal and screw-type orders are energetically degenerate with perpendicular orientations of the $\bm{Q}_{1,2}$-vector; $d_1$ and $J_{2}^{xy}$ lift this degeneracy.
The colour code indicates the ratio of the integrated cycloidal-type and screw-type spin structure factor (Supplementary Note \ref{Esec:Spin_Hamiltonian}) obtained for $B=0\,\mathrm{T}$ and $T=1\,\mathrm{K}$. 
With positive $J_{2}^{xy}$ and finite $d_{1}$, a helimagnetic cycloidal structure is favoured.
}
\label{FigS_MonteCarlo}
\end{figure}

\begin{figure}[h]%
\centering
\includegraphics[width=0.5\textwidth]{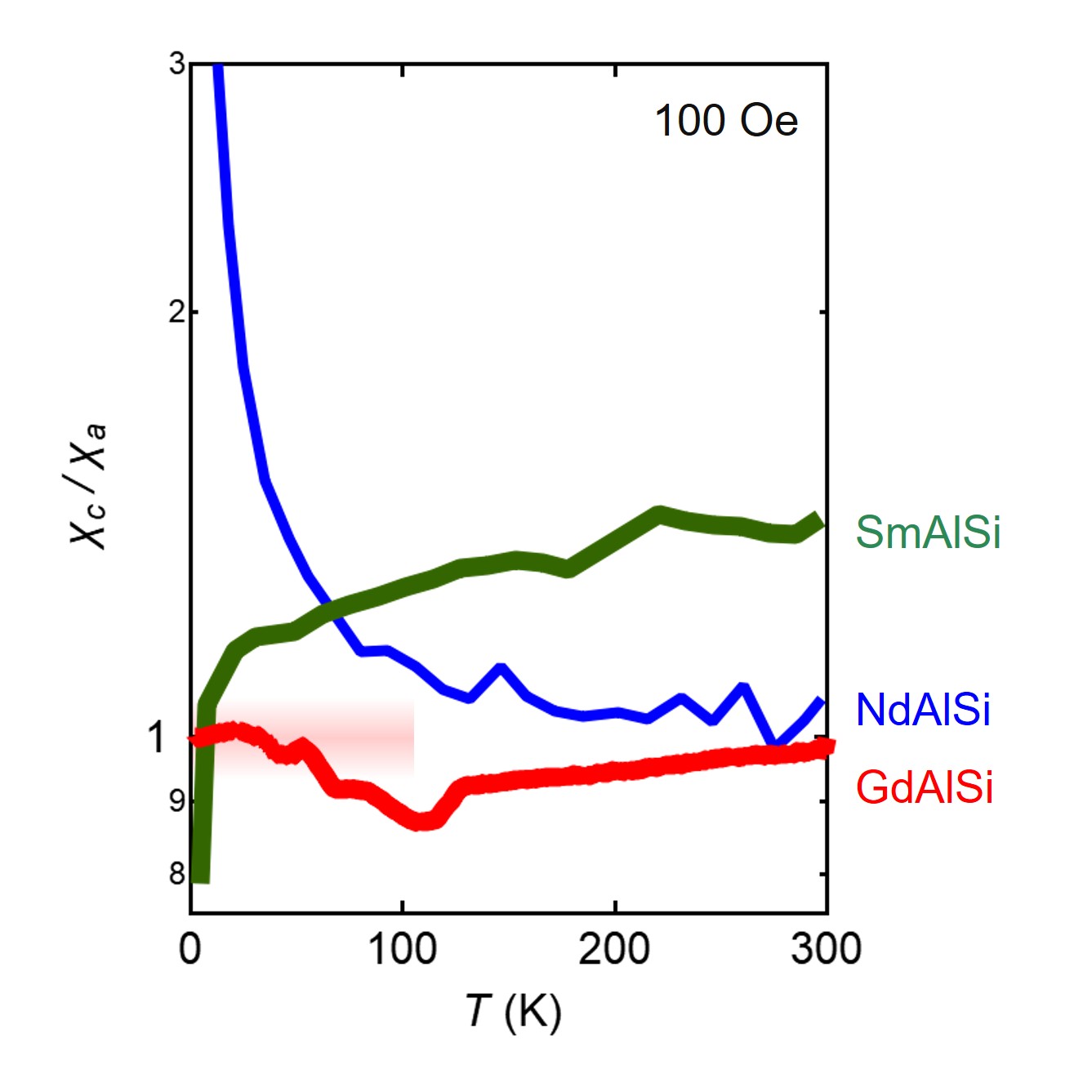}
\caption{
\textbf{Weak magnetic anisotropy in GdAlSi, as compared with NdAlSi and SmAlSi.}, Temperature ($T$-) dependence for the ratio of magnetic susceptibility for the out-of-plane field ($\chi_{c}$) and in-plane field ($\chi_{a}$). GdAlSi shows $\chi_{c}/\chi_{a}\simeq1$ at base temperature, highlighting extremely weak magnetocrystalline anisotropy consistent with helimagnetic order. NdAlSi and SmAlSi data are reproduced from Ref. \cite{yao_2023_SmAlSi}.
}
\label{FigS_aniso}
\end{figure}

\begin{figure}[h]%
\centering
\includegraphics[width=1.\textwidth]{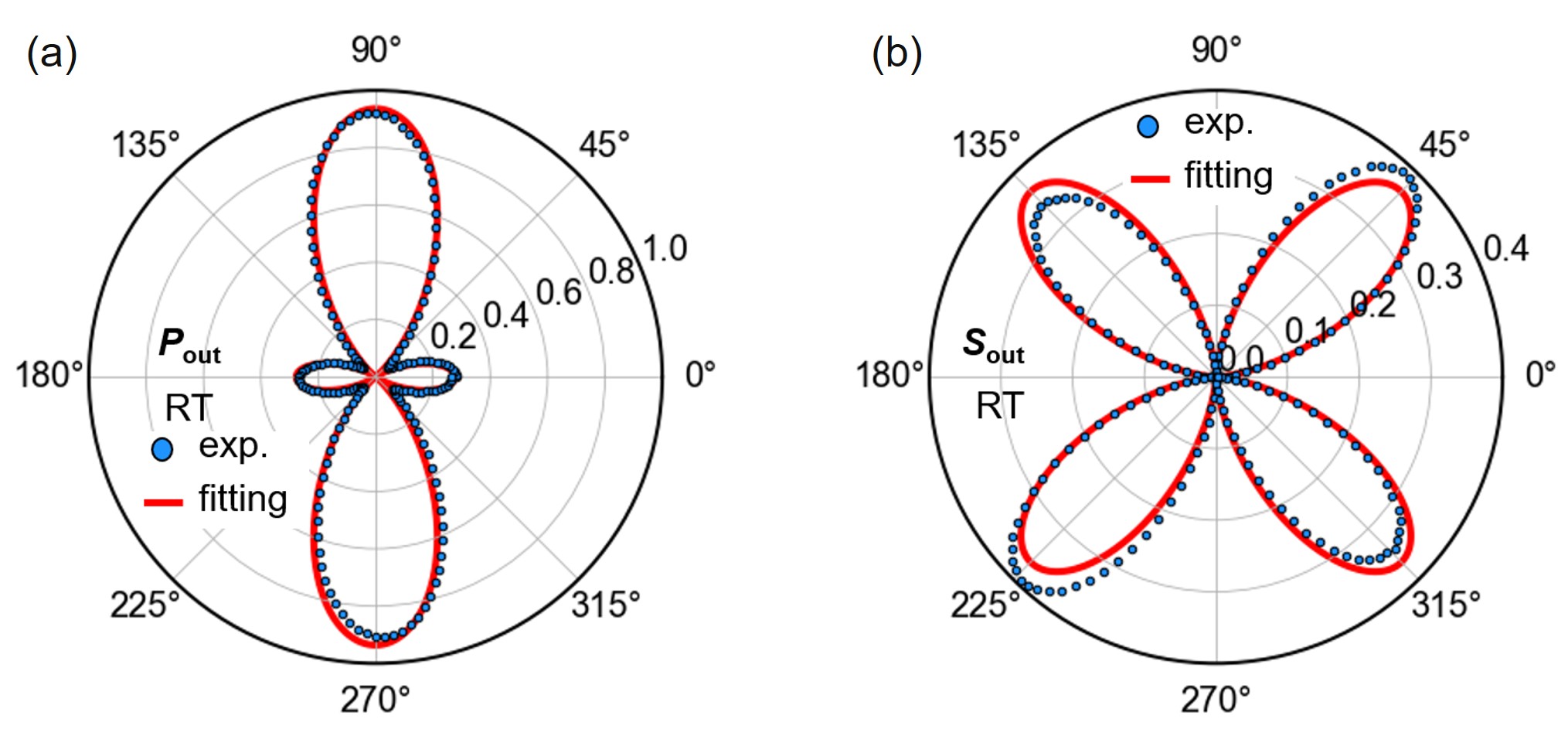}
\caption{
\textbf{Second harmonic generation (SHG) of light, evidencing the noncentrosymmetric crystal structure of GdAlSi.} (a,b) The horizontally polarized (a) and vertically polarized (b) SHG intensity when rotating the incident polarization angle at room temperature. The blue symbols and red curves represent the experimentally SHG responses and the fitting functions based on the symmetry argument, respectively (see Methods for details).
}
\label{FigS_SHG}
\end{figure}

\begin{figure}[h]%
\centering
\includegraphics[width=0.9\textwidth]{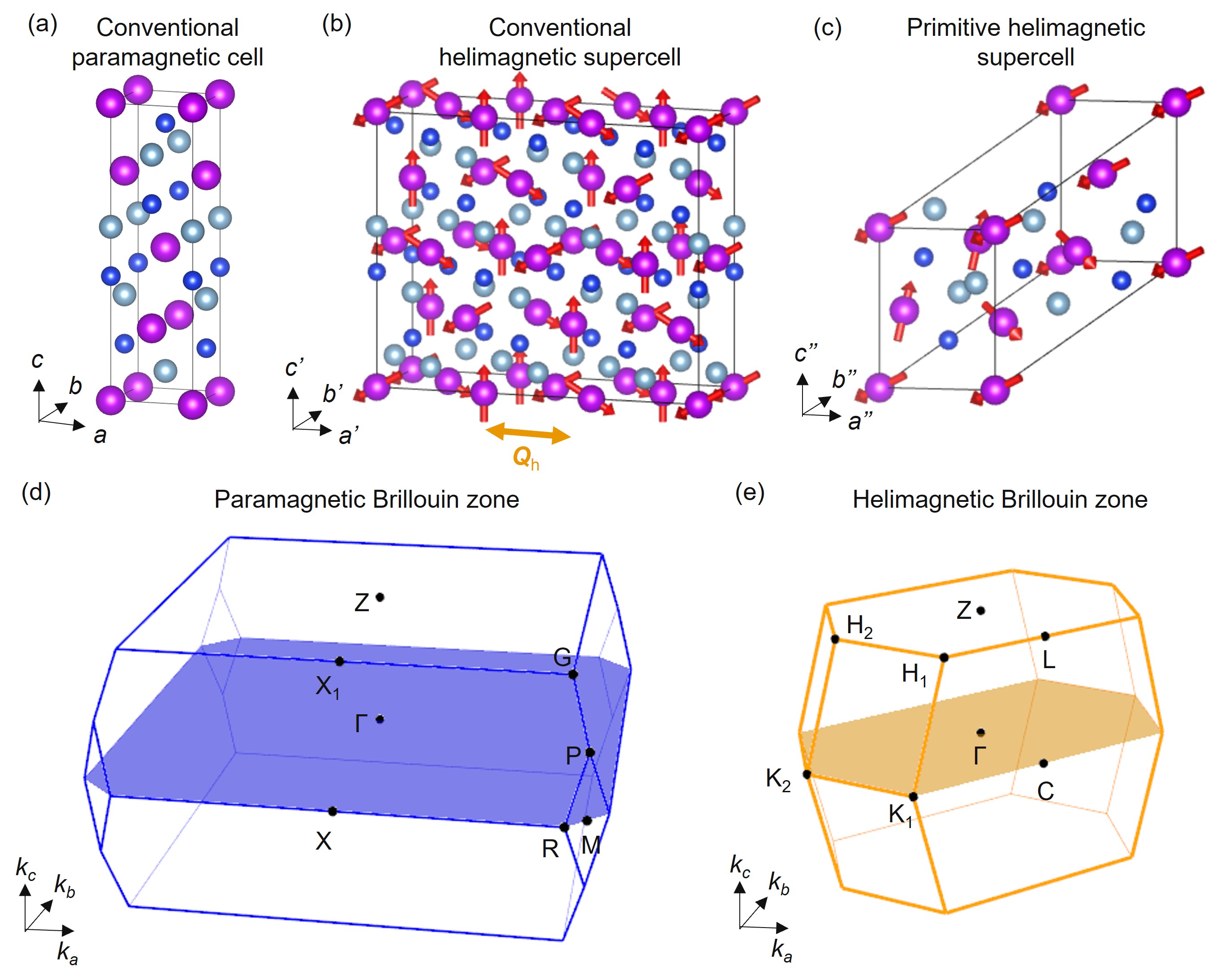}
\caption{\textbf{Crystallographic and magnetic unit cells in GdAlSi.} (a) Paramagnetic crystal structure of GdAlSi in the conventional cell representation. The lattice vectors are denoted as $\bm{a}, \bm{b},$ and $\bm{c}$. (b) Helimagnetic supercell with $\bm{Q}_\mathrm{h}= (2/3, 2/3, 0)$ in the conventional cell representation. The lattice vectors are $\bm{a}', \bm{b}',$ and $\bm{c}'$. The volume of the supercell is six times larger than that of the conventional paramagnetic cell. \textbf{c,} Helimagnetic supercell in the primitive cell representation. The lattice vectors are $\bm{a}'', \bm{b}'',$ and $\bm{c}''$. The volume of this primitive cell is 1.5 times larger than the conventional paramagnetic cell. (d,e) Corresponding paramagnetic (d) and helimagnetic (e) Brillouin zones. 
}
\label{FigS_unit_Brillouin}
\end{figure}

\clearpage

\section{Magnetic structure in $B = 7\,$T.}
GdAlSi experiences a transition from the harmonic helimagnetic cycloidal state to the distorted helimagnetic cycloidal state by applying an external magnetic field $\bm{B}\parallel[110]$. In this section, we report the detailed magnetic structure of the distorted helimagnetic cycloidal state by employing REXS measurements in a high magnetic field ($7\,\mathrm{T}$). In the applied field, the originally equivalent $[110]$ and $[1\overline{1}0]$ directions are no longer interchangeable. In the distorted cycloidal state of GdAlSi, the dominant scattering intensity ($> 80\,\%$) is along the $\bm{Q}_\mathrm{h}\perp \bm{B}$ direction; a weaker magnetic reflection appears at $\bm{Q}_\mathrm{uud}\parallel \bm{B}$. Thus, we investigate two types of magnetic reflections perpendicular ($\bm{Q}_{\mathrm{h}}\parallel[1\overline{1}0]$) and parallel ($\bm{Q}_{\mathrm{uud}}\parallel[110]$) to the magnetic field. 

To facilitate the discussion, the $\bm{Q}$-modulated magnetic moment $\bm{m}(\bm{Q})$ is separated into three mutually orthogonal Cartesian components, as in Eq. (\ref{eq_REXS_spin}) of the Main Text:
\begin{equation}
    \bm{m(\bm{Q})} = m_{c}(\bm{Q})\bm{e}_{c} + m_{110}(\bm{Q})\bm{e}_{110} +  m_{1\overline{1}0}(\bm{Q})\bm{e}_{1\overline{1}0}, 
\label{eq_REXS_spin_B}
\end{equation}
where $\bm{e}_{c}$, $\bm{e}_{110}$, and $\bm{e}_{1\overline{1}0}$ are unit vectors along the $[001]$, $[110]$, and $[1\overline{1}0]$ directions, respectively. 
Note that $m_{\mathrm{out}}(\bm{Q})$ in Methods is $m_{110}(\bm{Q})$, and $m_{\mathrm{in}}(\bm{Q})$ in Methods is separated into $m_{1\overline{1}0}(\bm{Q})$ and $m_{c}(\bm{Q})$.
We detect all three vector components of the modulated magnetization, by selecting two different reflections each for both $\bm{Q}_{\mathrm{h}}$ and $\bm{Q}_{\mathrm{uud}}$. For $\bm{Q}_{\mathrm{h}} = (q, -q, 0)$, we investigate the satellite magnetic reflections of the equivalent fundamental Bragg reflections $(-2, 2, 12)$ and $(2, -2, 12)$. Both reflections originate from the same magnetic domain.
As shown in Supplementary Fig. \ref{FigS_HF_spin}c,d, $\bm{m}(\bm{Q}_{\mathrm{h}})$ exhibits comparable $m_{c}(\bm{Q}_{\mathrm{h}})$ and $m_{1\overline{1}0}(\bm{Q}_{\mathrm{h}})$, but not $m_{110}(\bm{Q}_{\mathrm{h}})$, highlighting harmonic helimagnetic cycloidal behaviour as in the zero field state. For $\bm{Q}_{\mathrm{uud}} = (q, q, 0)$, two satellite magnetic reflections $(-1-q, 3-q, 10)$ and $(1+q, -3+q, 10)$ are investigated. They originate from the same magnetic domain. As shown in Supplementary Fig. \ref{FigS_HF_spin}e,f, $\bm{Q}_{\mathrm{uud}}$ is commensurate to the lattice with $q = 2/3$. Further, this magnetic wave exhibits only the $m_{110}(\bm{Q}_{\mathrm{uud}})$ Fourier component. We conclude that this subdominant modulation represents an up-up-down structure, where two moments align along $\bm{B}$, followed by one moment antiparallel to $\bm{B}$. 

\label{SNote_HF_spin}

\label{SNote_phase}

%
\section{Magnetic interactions and spin Hamiltonian in real space.}
\label{Esec:Spin_Hamiltonian} 
We perform Monte Carlo (MC) calculations using a real-space spin Hamiltonian. Our model truncates the Fourier-space model discussed in Methods and incorporates a minimal set of interactions to provide an intuitive physical picture of magnetic order in GdAlSi.

Firstly, the minimal model to explain the noncollinear magnetism in GdAlSi is based on only two frustrated exchange interactions: the in-plane nearest neighbour exchange $J_1$ and the inter-layer exchange $J_c$. The latter couples nearest neighbours along the $c$-axis as illustrated in Supplementary Fig. \ref{FigS_MonteCarlo}a.
The interaction Hamiltonian reads
\begin{equation}
    H_\text{ex} = 
    J_1 \sum_{\langle i,j \rangle_1} \hat{\mathbf{S}}_i \cdot \hat{\mathbf{S}}_j +
    J_c \sum_{\langle i,j \rangle_c} \hat{\mathbf{S}}_i \cdot \hat{\mathbf{S}}_j 
\end{equation}
where $\langle...\rangle_1$ and $\langle...\rangle_c$ are nearest in-plane and inter-plane neighbours, respectively, and $\hat{\mathbf{S}}_i$ is the normalized, classical Heisenberg spin (magnetic moment) on site $i$.
Consistent with the experiment, this natural choice of interaction terms (i) selects only propagation vectors of the type $\bm{Q}_{1,2}= (\pm q, \pm q, 0)$, constrained by tetragonal symmetry; (ii) reproduces the experimentally relevant range of $q$ for a reasonable $J_c/J_1$ ratio; (iii) demonstrates the competition between incommensurate helimagnetism and commensurate up-up-down order in $R$AlSi, $R = \,$rare earth. The red symbols in Supplementary Fig. \ref{FigS_MonteCarlo}c correspond to two subsets of simulations in finite magnetic field: those with $\bm{Q}_{1}$ perpendicular to the magnetic field (circles) and those with $\bm{Q}_{2}$ parallel to it (triangles). While the former dominate for incommensurate $q$, the latter only appear for commensurate $q=2/3$. The up-up-down order is well defined only at this unique value of $q$ and, here, it is energetically favoured over the helimagnetic state. 

Supplementary Fig. \ref{FigS_MonteCarlo}c thus demonstrates the fragile balance between incommensurate helimagnetism and commensurate up-up-down order in GdAlSi, consistent with discussion in Supplementary Note \ref{SNote_HF_spin}. This balance is sensitive to magnetic field, temperature, and the details of the exchange Hamiltonian. We highlight that the competition between the commensurate up-up-down and the incommensurate cycloidal order is realized simply from the $J_{c}/J_{1}$ dependence of $q$.

Next, inversion symmetry is broken on all interaction bonds in GdAlSi, leading to a variety of different \textit{antisymmetric} exchange interaction terms, characterized by Dzyaloshinskii-Moriya (DM) vectors favouring cycloidal-type or screw-type helimagnetism along specific bonds. 
However, all the DM interactions (DMI) which favour screw-type order come with oscillating sign between layers within the unit cell. Therefore, their effects cancel overall, when considering magnetic orders with translation symmetry along the $c$-axis.
The remaining DMI, which favours cycloidal order, is well approximated by an effective nearest neighbour DMI $d_1$, see Supplementary Fig. \ref{FigS_MonteCarlo}b.
Explicitly, the interaction reads
\begin{equation}
    H_\text{DM} = 
    d_1 \sum_{\langle i,j \rangle_1} \left( \hat{\mathbf{z}} \times \hat{\mathbf{r}}_{ij} \right) \cdot \left( \hat{\mathbf{S}}_i \times \hat{\mathbf{S}}_j \right)
\end{equation}
where $\hat{\mathbf{z}}$ is the unit vector along the $z$-direction (crystallographic $c$-direction) and $\hat{\mathbf{r}}_{ij}$ is the unit vector pointing from site $i$ to site $j$.
Intuitively, the effective DMI $d_1$ is understood to originate from mirror symmetry breaking by the ordered cage of Al and Si ions around each Gd ion.

Thirdly, anisotropic \textit{symmetric} exchange interactions are symmetry-allowed on all bonds:
In other words, the spin-components parallel to the interaction bond can have an interaction magnitude different from the other spin components.
For the $J_2$ bond, this anisotropic exchange interaction reads
\begin{equation}
    H_\text{anis} = 
    J_2^{xy} \sum_{\langle i,j \rangle_2} \left( \hat{\mathbf{S}}_i \cdot \hat{\mathbf{r}}_{ij} \right) \left( \hat{\mathbf{r}}_{ij} \cdot \hat{\mathbf{S}}_j \right)
\end{equation}
where $\langle...\rangle_2$ labels next-nearest neighbour pairs within the same plane, see Supplementary Fig. \ref{FigS_MonteCarlo}b. 
Note that the same kind of exchange anisotropy is allowed, in principle, for $J_1$ and $J_c$. However, $J_2^{xy}$ is the lowest-order term, when expanding in bond distance, which efficiently couples the spin components parallel to $[110]$ and parallel to $[1\bar{1}0]$, respectively. This term is consistent with the formation of the experimentally observed, distorted cycloidal order.
We further note that $90^\circ$ rotational symmetry about the bond direction is broken by the ordered Al-Si-cages, allowing for a different coupling strength of $S_z$ as compared to $S_x$, $S_y$. 
However, the experimentally observed cycloids are undistorted and harmonic within the experimental resolution, suggesting a secondary role for this $J_i^{zz}$ type of exchange anisotropy, where $i$ is a bond label.

Lastly, let the Zeeman interaction couple the spins to an external magnetic field $\mathbf{B}$ pointing along the $[110]$ direction.
In total, the energy functional used in our MC-simulations thus reads:
\begin{equation}
    H = H_\text{ex} 
    + H_\text{DM} 
    + H_\text{anis} 
    - \mu_B\, S\, \mathbf{B}\cdot \sum_i \hat{\mathbf{S}}
\end{equation}
with $S=7$ being the total spin of the $4f$ electron shell of the Gd-ion.
This energy functional is implemented in our self-written GPU-accelerated Metropolis-Monte-Carlo-simulator.
The resulting competition between DMI and anisotropic exchange is depicted in Supplementary Fig. \ref{FigS_MonteCarlo}d, where DMI favours cycloids and anisotropic exchange stabilizes either cycloid or screw, depending on its sign. Both types of magnetic order can be realized depending on the relative strength of these interactions.

Our real-space Hamiltonian model, based on the reciprocal space theory of Eq. (\ref{eq_Kondo}) in Methods, therefore, demonstrates that the $R$AlSi family ($R = \,$rare earth) can realize competition among commensurate up-up-down and incommensurate cycloidal / screw orders. In addition to our present observations for GdAlSi, recent neutron scattering works on NdAlSi (commensurate up-up-down) and SmAlSi (incommensurate cycloid or screw) seem to also support this notion~\cite{gaudet_2021_NdAlSi,yao_2023_SmAlSi}.

\end{document}